# WAP: Digital Dependability Identities

Daniel Schneider,
Mario Trapp
Fraunhofer IESE
Kaiserslautern, Germany

Yiannis Papadopoulos
University of Hull,
Hull, United Kingdom

Eric Armengaud
AVL List
Graz, Austria

Marc Zeller, Kai Höfig
Siemens AG
München, Germany

*Abstract*—Cyber-Physical Systems (CPS) provide enormous potential for innovation but a precondition for this is that the issue of dependability has been addressed. This paper presents the concept of a Digital Dependability Identity (DDI) of a component or system as foundation for assuring the dependability of CPS. A DDI is an analyzable and potentially executable model of information about the dependability of a component or system. We argue that DDIs must fulfill a number of properties including being universally useful across supply chains, enabling off-line certification of systems where possible, and providing capabilities for in-field certification of safety of CPS. In this paper, we focus on system safety as one integral part of dependability and as a practical demonstration of the concept, we present an initial implementation of DDIs in the form of Conditional Safety Certificates (also known as ConSerts). We explain ConSerts and their practical operationalization based on an illustrative example.

## I. Introduction

Cyber-Physical Systems (CPS) harbor the potential for vast economic and societal impact in domains such as mobility, home automation and delivery of health. At the same time, if such systems fail they may harm people and lead to temporary collapse of important infrastructures with catastrophic results for industry and society. Thus, ensuring the dependability of Cyber Physical Systems is the key to unlocking their full potential and enabling industries to develop confidently business models that will nurture their societal uptake. Using currently available approaches, however, it is generally infeasible to assure the dependability of Cyber-Physical Systems. CPS are typically loosely connected and come together as temporary configurations of smaller systems which dissolve and give place to other configurations. The key problem in assessing the dependability of CPS is that the configurations a CPS may assume over its lifetime are unknown and potentially infinite. State-of-the-art dependability analysis techniques are currently applied during design phase and require a priori knowledge of the configurations that provide the basis of the analysis of systems. Such techniques are not directly applicable, can limit runtime flexibility, and cannot scale up to CPS.

This paper addresses this important and unsolved problem by presenting the novel concept of a Digital Dependability Identity (DDI) for (1) improving the efficiency of generating consistent dependability argumentation over the supply chain during de-sign time, and (2) laying the foundation for runtime certification of ad-hoc networks of embedded-systems. A DDI is an analyzable, composable and potentially executable model of information about the dependability of a component or system which is maintained (and evolved) over their complete lifecycle. Note, however, that even though we introduce and discuss DDI with the broader scope of dependability, we exemplify it only for safety.

The paper is structured as follows. In Sec. 2, we discuss current industrial and societal needs which motivate this research. We focus on the automotive, railway and, later in the example, the agriculture domains and show that these challenges exist in similar form in other domains. We argue that DDIs must fulfill a number of properties including: a) be usable across supply chains b) provide means for off-line certification of systems, and, c) enable public checks of compliance of components and systems with the requirements of a CPS and provide capabilities for run-time certification of safety of CPS in-the-field. In Sec. 3, we outline the general concept of a DDI and discuss how it can address the identified requirements. In Sec. 4, we present an initial safety-centered implementation of DDIs in the form of Conditional Safety Certificates (ConSerts). In Sec. 5 we discuss related work, and, finally, in Sec. 6 we conclude and point towards future work.

## II. Technological Challenges & Requirements for DDIs

CPS are emerging in many industries. In the automotive industry the market is strongly driven by competition and influenced by variety of legislation, e.g. regarding passenger car emissions targeting the reduction of $CO^2$ emissions, and by directives, e.g., to reduce the number of fatalities on the road. Current major trends include the introduction of embedded systems for passive safety (protection during crash) and active safety (crash prevention). Examples are anti-lock braking systems (ABS), airbags, electronic stability control (ESP) or advanced driver assistance systems (ADAS) such as adaptive cruise control, lane departure warning systems, etc. An important aspect for ADAS are the increasing levels of autonomy envisioned for CPS which anticipate situations in which the vehicle performs all safety-critical functions for the entire trip, with the driver not expected to control the vehicle at any time. A key enabler for the development of ADAS is the capability of the vehicle to correctly apprehend its environment – possibly by communicating with its surroundings (e.g. connected vehicles), and to take safety-critical decisions in a cooperative manner during field-operation and between different systems from different manufacturer. The railway domain experiences similar challenges with heterogeneous systems of systems that adhere to different standards and system qualities. For example, the European Train Control System (ETCS) provides standardized train control in Europe and eases travelling with trains crossing the borders of all European countries. Nevertheless, local train control systems are still in use and are also capable to be instrumented for ETCS. Since not only different trains are produced, but also different track side solutions are present even within one country, certification activities (w.r.t. CENELEC EN 50126 & 50129) require accurate planning and must react quickly on changes within system development. Systems of systems are also produced by various stakeholders in the value chain and, therefore, safety information about components and subsystems (rolling stock, track-side and railway systems) need to be interoperable. The dependability assurance of such systems poses significant challenges given that dependability assurance cases for complex systems must be devised from information about modules that are independently developed across the entire value chain. To address such challenges in complex CPS or systems of systems, we develop the concept of a DDI, i.e. a modular, composable and potentially executable, dependability specification. In the remainder of the section we identify key requirements for DDIs.

*A. Universal exchange of dependability information*

Like the systems that compose a CPS, we expect that DDIs will be produced by multiple stakeholders in a supply chain and, therefore, they need to be interoperable and expressed in a common communication language that can be understood by stakeholders and mechanisms undertaking the generation and evaluation of DDIs. Although progress has been achieved with dependability meta-models, e.g. within architecture description languages like EAST-ADL and AADL, there is still a lack of a common model for the communication of dependability information-. This is needed in industrial practice, where it is unrealistic to expect that all companies along a value chain should be forced to use a unified methodology. For this reason, a precondition for DDIs is the existence of an open dependability meta-model, which can provide the basis for expressing DDIs. This meta-model should be independent from specific development approaches and tools and enables collaboration between actors in the value chain. For DDIs to work as medium for synthesis of heterogeneous dependability information, they must be sufficiently expressive to enable the component integrator to compile DDIs from the DDIs of subcomponents. Moreover, DDIs should optionally abstract from details to protect the component provider's IP.

*B. Efficient dependability assurance across industries and value chains*

In the current industrial practice, different dependability tools and methodologies are used. For this reason, it must be possible for a component provider to generate DDIs based on the information that is already available in the tools that are established in a company. Moreover, it must be possible to include the information contained in DDIs into the established dependability assurance lifecycle and tool chain of the component integrator. In a typical scenario, system and component requirements and their integration context change constantly; these changes need to be reflected in the dependability analyses of a system. This in turn requires the ability to perform semi-automated change-impact analyses across the DDIs of components to reflect design changes in a system DDI. As a further aspect, it is typical for companies to develop product families instead of single products. For example, in the automotive industry, it is not unusual to assure the safety of more than a thousand variants of a powertrain control system per year. This requires that DDIs support variability, including analyses that enable a prediction about whether a component will fit into multiple product variants, to help reduce re-assurance effort.

Finally, there is the challenge on how to control dependability from the early stages of design, so that it is not treated as emergent property. This is important for the synthesis of DDIs, because controlled processes with rational allocation of requirements that can work across the tiers of a value chain can also assist the effective collection and synthesis of DDIs. Numerous safety standards in various industries, such as IEC61508, ISO26262 or ARP4754, envisage processes of refinement which are driven by safety requirements expressed as Safety Integrity Levels (or Development Assurance Levels). These processes implicitly define similar patterns for the construction of dependability assurance cases that can be captured and their model-based synthesis can then become possible.

*C. Dependable integration of systems in the field*

In CPS, dependability cannot be fully assured prior to deployment. Indeed, systems will dynamically interconnect and form systems of systems with largely unpredictable consequences for dependability. In order to assure the dependability of such in-field integrations, the degree of automation in the evaluation of DDIs must be further increased to include additional forms of runtime or in-the-field evaluation that will enable or disable operations in a particular configuration or context. DDIs therefore must become executable specifications, not simply digital artifacts that cease in their utility after deployment of the system. DDIs must be made publicly available, to all companies whose systems shall be integrated into a particular CPS, so that developers of new systems can check whether their system can be integrated in a dependable way. Companies shall have the possibility to stay informed about changes in CPS in order to check if this requires modifications in their systems. For highly dynamic environments, it is additionally necessary to enable a fully automated evaluation of DDIs so that it is possible to decide, without human interaction if needed, whether or not a dependable collaboration is possible. Such a check based on the system DDIs should be possible off-board (e.g., in the cloud) as well as on-board, in the latter case with the DDIs as executable specifications and being evaluated by the systems themselves.

III. DIGITAL DEPENDABILITY IDENTITIES

In general, a Digital Identity is defined as "the data that uniquely describes a person or a thing and contains information about the subject's relationships" [2]. Applying this idea, a DDI contains all the information that uniquely describes the dependability characteristics of a system or component. This includes attributes that describe the system's or component's dependability behavior, such as fault propagations, as well as requirements on how the component interacts with other entities in a dependable way and the level of trust and assurance, respectively. The latter can be described using concepts from the theory of safety contracts. A DDI is a living dependability assurance case. It contains an expression of dependability requirements for the respective component or system, arguments of how these requirements are met, and evidence in the form of safety analysis artifacts that substantiate arguments. A DDI is produced during design, issued when the component is released, and is then continually maintained over the complete lifetime of a component or system. DDIs are used for the integration of components to systems during development as well as for the dynamic integration of systems to "systems of systems" in the field. For their realization, DDIs require three components which are currently under development.

### A. An Open Metamodel for specifying DDIs

This is an Open Dependability Exchange (ODE) meta-model enabling exchange and integrated analysis of modular yet heterogeneous dependability-related information over supply chains. This ODE meta-model provides means to specify and connect dependability information, like hazard analyses, failure propagation models and safety argumentation blocks, into coherent modular safety arguments about components or systems. It is also possible to specify the level of trust for this argumentation with respect to trust on the issuer and to the trust level of the promised services during field operation. This information is represented independently from a concrete development approach. The ODE can represent typical models that are available in dependability assurance lifecycles in an abstract fashion. Therefore, it is possible to transform available models like fault trees or existing safety cases to ODE format.

### B. Means for off-line synthesis and evaluation of DDIs

Means for (semi-)automated synthesis and evaluation of DDIs are being provided. DDIs can be created from existing dependability information. To this end, it is feasible to transform the information stored in existing tools like FaultTree+ into an ODE-compliant model. Moreover, semi-automated algorithms generate DDIs based on modular DDI specifications expressed in the ODE-compliant model. This is achieved using mechanisms for abstraction, simplification, information hiding and formalization. Once DDIs of components are available, a component integrator is able to include them in a system DDI. To this end, it is important to provide semi-automated change impact analyses supporting the developers in the handling of modifications in DDIs during the lifecycle. To enable a further level of automation, we use state-of-the-art technologies for model-based dependability analysis to develop a system for fully automated synthesis of DDIs. This system creates DDI structures from systematic refinements of system architectures according to the dependability requirements allocation processes and corresponding algebras described in modern dependability standards like ISO26262. The goal structures that form the skeleton of the DDI of a system is supplemented with evidence of dependability in the form of fault trees and FMEAs that are created by means of automated model-based dependability analyses. These methods are built on a cutting edge algorithm for automatic allocation of dependability requirements that exploits meta-heuristics and techniques for model-based dependability analysis [3]. Assuming that a system has its dependability requirements expressed in the form of integrity levels and has a proposed architecture for its implementation, the algorithm calculates optimal integrity level allocations, taking into account their dependencies and assumptions about their intended failure behavior. Stakeholders in a value chain will be able to apply this tool iteratively in order to specify safety requirements to suppliers in lower tiers. The algorithm will guarantee that a system will meet its dependability requirements at the end of the design refinement process if the basic components of the architecture also meet their dependability requirements; assuming that any assumptions of dependence and independence made in the model have not been violated. The decomposition itself together with progressive refinement of models will form the basis for the automated synthesis of DDIs.

### C. Means for in-the-field evaluation of DDIs

While the integration of components to systems follows traditional value chains with contractor-supplier relationships, the dynamic integration of systems of systems is rather a collaboration of equal partners. In order to enable the dependable integration of CPS, additional concepts are required. First, it is necessary that the DDIs of a CPS that is already in the field are made available in a central registry, while protecting intellectual property rights. Based on the centrally available DDIs, a system manufacturer can check whether or not his system can be dependably integrated with the already existing systems of the CPS. In highly dynamic integrations, it is necessary to enable the onboard evaluation of DDIs. To this end, every system must store its DDI onboard and must be equipped with fully automated onboard evaluation algorithms, which enable the systems to determine whether they can collaborate dependably with other systems they want to connect to. As a result of the evaluation, the level of degradation to be applied for safe operation of the CPS shall be identified. One possible engine for in-the-field evaluation of DDIs that we currently evaluate is Complex Event Processing (CEP) [4]. CEP allows the identification and analysis of events, and the responding actions, to be made in real time. We use DDI specifications as rule bases that inform the in-the-field evaluation of DDIs by a CEP engine. DDIs model event patterns and cause-effect relationships between failures and effects that can be used for detection and diagnosis of events and event patterns by the CEP engine. The dependability with which the services of a component can be provided to consumers of those services can then be established.

The concept of DDI outlined so far is generic and can be realized in various ways. Next, we present Conditional Safety Certificates (ConSerts) as a first step towards DDI.

## IV. CONSERTS: A FIRST STEP TOWARDS DDIS

Conditional Safety Certificates (ConSerts) [5] are focused on one of the most challenging aspects of DDIs, the runtime integration scenario. In this regard, DDIs act as machine-readable modular dependability specifications that can be composed and analyzed between the systems in order to come to a dependability assessment of the resulting system composition. By providing this capability ConSerts are an initial "embryonic" implementation of DDI: they exhibit some characteristic features of a DDI, but they are technically not sufficient for all envisioned DDI use cases. They are focused on safety instead of all properties related to dependability and, as of yet, they lack the required maturity.

### A. Conditional Safety Certificates

ConSerts operate on the level of safety requirements. They are issued at development time and certify specific safety guarantees that depend on the fulfillment of specific demands regarding the environment. In the same way as "static" certificates, ConSerts shall be issued by safety experts, independent organizations, or authorized bodies after a stringent manual check of the safety argument. To this end, it is mandatory to prove all claims regarding the fulfillment of provided safety guarantees by means of suitable evidence and to provide adequate documentation of the overall argument – including the external demands and their implications.

There are some significant differences between ConSerts and static certificates that are owed to the nature of open systems: A ConSert is not static but conditional; it therefore comprises a number of variants that are conditional with respect to the (dynamic) fulfillment of demands; and it must be available in an executable (and composable) form at runtime. Conditions within a ConSert manifest in relations between potentially guaranteed safety requirements, that can simply be denoted as guarantees, and the corresponding demanded safety requirements, i.e. demands. Demands always represent safety requirements relating to the environment of a component, which cannot be verified at development time because the required information is not available yet. These demands might directly relate to required functionalities from other components. On the other hand, evidence can be required beyond that, since safety is not a purely modular property and it cannot be assumed that a composition of safe components is automatically safe. To this end, ConSerts support the concept of so-called Runtime Evidences (RtE) as an additional operand of the conditions. RtEs are a very flexible concept. In principle, any runtime analysis providing a Boolean result can be used. RtEs might relate to properties of the composition or to any context information, e.g. a physical phenomenon such as the temperature of the environment that is safety relevant. For example, RtE could relate to a physical phenomenon such as the temperature of the environment which could be measured with a sensor. Other RtE require dynamic negotiation between components.

In any case, ConSerts must be available at runtime in a machine-readable representation and the systems need to possess mechanisms for composing and analyzing runtime models. Based thereon, a valid safety certificate for the overall system of systems can be established.

### B. Example

We present ConSerts with an example from the agricultural domain which today pioneers innovative applications involving systems of systems and dynamic integration. One of these applications is the so-called *Tractor Implement Management (TIM)*. The TIM functionality enables agricultural implements to control the typical tractor functions such as velocity, steering angle, power take off, or auxiliary valves. It is possible to fully automate implement-specific work procedures and to optimize them with respect to parameters such as performance, efficiency, or wear and tear. TIM utilizes a standardized bus system for communication between the participating devices and machines. During TIM operation, control is typically assumed by the implement ECU. It uses the TIM functions of the tractor and devices such as sensors for the respective automation purpose, displays data to the operator, and executes operator inputs. Between different tractors, implements and auxiliary devices such as virtual terminals (providing the operator UI) or GPS systems of different manufacturers, a huge space of configurations arises which makes it unfeasible to analyze each potential configuration a priori at development time. For this reason, those TIM applications already available on the market today only work for prefixed concrete pairs of tractors and implements, whose integration has been thoroughly analyzed at development time by the involved manufacturers.

Assume there is a farmer who owns a TIM-capable tractor and a TIM-capable round baler. The TIM baling application is running on the implement ECU, the user interface is displayed on a virtual terminal in the cabin. In addition to a standard configuration, the baling application also supports an extended configuration that additionally incorporates a swath scanner device. This device is mounted in the front of the tractor and measures the volumetric flow and the location of the swath to further optimize the baling operation in terms of tractor speed and steering angle. The baling application can be enabled when tractor, implement, virtual terminal and swath scanner are connected and the ConSert-based interoperability and safety checks have been successful. Corresponding information is provided to the operator via the terminal. The actual round baling process can then be activated by the operator, who thus relinquishes control to the round baler. The round baler commands the tractor to drive over the swath with optimal acceleration rates and speed. When the bale reaches a preset size, the tractor decelerates to standstill and the bale is ejected. The process can then be re-started by the operator.

### C. Engineering of ConSerts

For the engineering of ConSerts in this example the role of the implement manufacturer shall be assumed. The goal of the manufacturer is to develop a round baler with TIM support. From a functional point of view, it is clear (due to existing standards) how the interfaces between the potential participants look like and how they are to be used. However, the implement manufacturer does not know about the safety properties of these functions.

From a safety point of view, the engineering of the baling application starts top-down with an application-level hazard and risk analysis. Assume that the agricultural manufacturers agreed by convention that during the operation of a TIM application, the application (and thus the application manufacturer) has the responsibility for the overall automated system. Therefore, the safety engineering goal is to ensure adequate safety not only for the TIM baling application or for the implement, but for the whole collaboration of systems that will be rendering the application service at runtime. Thanks to the ConSert-based modularization it is thereby sufficient to only consider the direct dependencies of the system under development on its environment. Potential "external" safety requirements will be either associated with demands regarding required services or with RtEs. At runtime, it will be determined whether the demands can be satisfied based on guarantees given by external systems. This negotiation can obviously range across several layers and incorporate series of dependent systems and guarantee-demand relationships.

Correspondingly, relevant hazards of the TIM baling application might be a self-acceleration or self-steering during operation or a self-acceleration or a power take of commission during standstill. These hazards would be assessed with respect to their associated risks based on the risk assessment tables provided by the ISO25119 (i.e. the safety standard of the agricultural domain). In a subsequent step, corresponding top-level safety requirements would be derived. In addition, reasonable guarantee levels are to be identified. In the given example it is conceivable that different guarantee levels are required for different locations (e.g. in the midst of nothing vs. a field close to a playground for children) or that guarantee levels are defined in interplay with application specific parameters (e.g. acceleration or velocity levels, different degrees of automation, etc.).

The next step is to develop a safety concept that ensures the satisfaction of the safety requirements and of the associated ConSert guarantees. This is done in a standard way: Safety analyses are applied to identify cause-effect relationships and to specify the failure logic, corresponding safety measures are identified and eventually, a conclusive safety argument is build up factoring in suitable evidence. A difference to the engineering of closed systems is that besides possible internal causes, there might be external causes that may either be associated with safety properties of the required services or with RtEs. Moreover, there is also some degree of variability to be considered due to different ConSert guarantees and corresponding differences in the correlated demands.

With regard to the causes related to required services, there are two possibilities. First, it is possible to define internal measures, such as error detection mechanisms, so that failures of the required services can be tolerated. Alternatively or in addition, it is possible to demand that the external service provider has to guarantee certain safety properties for the service. These safety properties need to be formalized and standardized for a domain in order to constitute the basis for the definition of ConSerts guarantees and demands. As for the RtEs, two categories can be distinguished: intra-device and inter-device RtEs. The former can be designed and implemented rather freely because they do not require any information from other external systems. The latter do require such information and thus, they need to be standardized or at least described in guidelines for a given domain. In reference to the example, assume that there is a top-level safety requirement that a *self-acceleration must not occur during standstill*. Based on the hazard and risk analyses it has been determined that this requirement needs to be assigned with *AgPL d[1]*. However, this is due to a relatively high exposure that is assumed for bystanders, as it would be the case for operation in areas close to housing. In other areas, *AgPL c* would be sufficient[2]. With ConSerts it is now conceivable to optimize the trade-off between availability and safety by factoring in dynamic context knowledge. In concrete this means that it would be possible to use, for instance, a GPS position and (in this case) annotated map data to distinguish between different usage contexts of the TIM system that imply different levels of safety requirements. Of course, such a mechanism needs to be safe in its own right, but for now let us just assume this can be done. Thus, three different levels of ConSert guarantees are defined in the example: a) a high integrity one, enabling full automation features of the TIM application b) a medium integrity one, enabling full features only in specific areas (or, alternatively, enables operation with some constraints) and c) a default guarantee that can always be granted, enabling only a very constrained operation, e.g. without acceleration from standstill or automated steering.

The high integrity guarantee would include *AgPL d* for self-acceleration in standstill as well as a series of other relevant guarantees omitted here for simplicity. The specification of the guarantee given next is based on a grammar and on service types, safety property types and rules of refinement specified in a domain-specific standard or guidelines:

```
TIMBalingSwSc(1): AgPL = b,
SelfAcc{,Standstill}.AgPL = d,
LateAcc{30s,Standstill}.AgPl = d, (...)
```

The first element of the specification denominates the associated service (by type) and gives an (absolute) order number for the guarantee (from 1 (best) to n (worst)). More sophisticated orderings could be useful but are not yet developed. The next element describes an integrity level for the whole service. This is basically a shortcut and implies that all safety properties of the service (as specified by the standard or guideline) are guaranteed with the named integrity level. Then a series of concrete safety properties is following, whose types and refinement parameters are also given by the standard.

The next step from a ConSert perspective is to determine the demands (i.e. service related demands as well as RtEs) that relate to the identified guarantees. This relation is modeled by means of a Boolean function, where the demands are input variables and the guarantee is the output variable. There is also a corresponding graphical specification technique based on directed acyclic graphs, where each function is represented by a tuple (D, R, BG, E, g): A set of Boolean input variables D representing service-related demands and RtEs R, a set of Boolean gates BG, a set of directed edges E connecting the elements, and, a Boolean output variable g.

Overall, a ConSert is a set of such functions, one for each guarantee level (of each offered service). The definition of the demands and the functions is done together with the development of the safety concept and safety argument. In fact, the resulting ConSert becomes an integral part of the safety argument, because it needs to be shown in a convincing manner that the ConSert guarantees are actually valid given the fulfillment of their related demands.

---

[1] E.g. controllability 3 (non-controllable), severity 3 (life threatening) and exposure 3 (often; 1-10% of operating time). As per ISO 25119.

[2] E.g. controllability 3, severity 3 and exposure 3 (sometimes; 0.1-1% of the operating time). As per ISO 25119.

*D. ConSerts at Runtime*

ConSerts need to be transferred into a machine-readable form to enable dynamic evaluation and there need to be corresponding mechanisms build into the systems that operate on this information to conduct the evaluation. Of course, the evaluation protocols need to be standardized to ensure that every participating system is interoperable from a ConSert point of view. Assume that the operator has installed the swath scanner on the tractor and that tractor and round baler are coupled. The operator initiates TIM via the virtual terminal and explicitly selects the application service variant that provides flexible control of speed and steering based on the input from the swath scanner. The first step is now to establish the application, i.e., to dynamically integrate the participating systems. After this has succeeded, the evaluation of safety guarantees of the application service is started. Note that the application service forms the root of a dynamically formed composition hierarchy and the correlated ConSert has the scope of this whole system of systems application. The evaluation of ConSerts starts from the leaf systems that have no external service-related dependencies. These systems determine their RtEs and propagate them up in the composition hierarchy. Eventually, all service-related demands of the root (i.e. the TIM baling application) can be checked and together with the evaluation results of the RtEs the top level safety guarantees are determined.

## V. RELATED WORK

Modular, model-based safety analysis has been developed since the 1990s. Examples are the Failure Propagation and Transformation Notation (FPTN) [6], HiP-HOPS (Hierarchically Performed Hazard Origin and Propagation Studies) [1], and Component Fault Trees (CFTs) [7]. In terms of safety assurance rather than safety analysis, the key concept of the past two decades has been the development of the safety case. Over time, methods to support the creation of safety cases have been developed, e.g. the Safety Concept Tree approach [8] and the Goal Structured Notation (GSN) [9]. Modularization came up as an important topic which is especially relevant for the concept presented in this paper. Notable work in this regard has been done by Kelly et al.(e.g. the modular GSN [9]), Gallina et al. (e.g. contract-supported argument fragments [15]), and, recently by Denney et al. (dynamic safety cases [16]). Other modular safety assurance and certification approaches include the "Open Certification" model [10] in the context of IEC 61508, and the approach on incremental safety assurance as part of a bottom-up certification process in the aerospace domain [11]. Further relevant work in this area includes the work on the modularization of safety concepts as part of the DECOS (Dependable Embedded Components and Systems) project [12], which balances reusable, generic safety cases and application-specific safety cases, as well as the work on the VerSaI requirements specification language [13]. Initial ideas for runtime certification based on formal analyses, enabling the verification of component runtime behavior according to its specification are discussed in [14].

## VI. CONCLUSION

In this paper, we started from the observation that current approaches to certification are generally limited to design-time safety assurance. Such approaches struggle to adequately assure the safety of dynamic Cyber-Physical Systems whose configurations may change in the field as they adapt to their environments or as other systems join or leave. Modular safety assurance techniques provide an important starting point, but for them to function in such a context, they need to be adapted to the demands of runtime field operation. To that end, we developed the novel concept of Digital Dependability Identities as a framework for dependability assurance of CPS that spans the lifecycle and is usable in a complex and realistic procurement environment. The requirements for DDIs were elicited and form the main conceptual contribution in section 3. Conditional Safety Certificates were also presented as a constrained embryonic implementation of DDIs to demonstrate key functionalities of DDIs, in particular the concept of runtime certification using executable dependability analyses. We are currently expanding this work on ConSerts to provide the full specification of DDIs outlined in section 3. Particular focus will thereby be laid on the interplay between safety and security, since vulnerabilities in a CPS context can easily become safety issues. One further aspect of focus is resolving issues of partial observability and uncertainty in the run-time evaluation of ConSerts and DDIs.